\documentclass[12pt,fleqn,twoside]{article}
\usepackage{amsfonts}

\def\noi{\noindent}

\makeatletter
\renewcommand{\section}{\@startsection{section}{1}{0pt}%
        {-3.5ex plus -1ex minus -.2ex}{2.3ex plus .2ex}%
        {\large\bf\protect\raggedright}}

\renewcommand{\subsection}{\@startsection{subsection}{2}{0pt}%
        {-3ex plus -1ex minus -.2ex}{1.4ex plus .2ex}%
        {\normalsize\bf\protect\raggedright}}

\renewcommand{\@oddhead}{\raisebox{0pt}[\headheight][0pt]{%
   \vbox{\hbox to\textwidth{\rightmark \hfil \rm \thepage \strut}\hrule}}}
\renewcommand{\@evenhead}{\raisebox{0pt}[\headheight][0pt]{%
   \vbox{\hbox to\textwidth{\thepage \hfil \leftmark \strut}\hrule}}}
\newcommand{\heads}[2]{\markboth{\protect\small\it #1}{\protect\small\it #2}}

\makeatother

\newcommand{\Title}[1]{\noi {\uppercase{\Large #1}} \\}
\newcommand{\Author}[2]{\noi{\large\bf #1}\\[2ex]\noindent{\it #2}\\}
\newcommand{\Abstract}[1]{\vskip 2mm \begin{center}
        \parbox{16.4cm}{\small\noi #1} \end{center}\medskip}

\def\nq{\hspace{-1em}}                      \def\cm{\hspace*{1cm}}
\def\nqq{\hspace{-2em}}                     \def\inch{\hspace*{1in}}
\def\nhq{\hspace{-0.5em}}                   
                   \def\sect{Sec.\,}

      \def\beq{\begin{equation}}
\def\eq{Eq.\,}                              \def\eeq{\end{equation}}
\def\eqs{Eqs.\,}                            \def\bear{\begin{eqnarray}}

\def\al{&\nhq}                              \def\lal{&&\nqq {}}
\def\bearr{\begin{eqnarray} \lal}           
\def\ear{\end{eqnarray}}                    
\def\earn{\nonumber \end{eqnarray}}         \def\nnn{\nonumber\\ \lal }
                        \def\nnnv{\nonumber\\[5pt] \lal }
\def\dst{\displaystyle}                     \def\yy{\\[5pt]}
 \def\yyy{\\[5pt] \lal}
 \def\eql{\al =\al}

      \def\const{{\rm const}}
\def\e{{\,\rm e}}                      \def\Half{{\dst\frac{1}{2}}}
\def\d{\partial}

\def\sign{\mathop{\rm sign}\nolimits}  \def\ten#1{\mbox{$\cdot 10^{#1}$}}
  \def\dim{\mathop{\rm dim}\nolimits}

\newcommand{\vars}[1]{\left\{\begin{array}{ll}#1\end{array}\right.}

\def\ep{\epsilon}

\def\mn{_{\mu\nu}}          \def\od{{\overline d}{}}
\def\MN{^{\mu\nu}}          
           \def\og{{\overline g}{}}
           
                            \def\ocR{\overline{\cal R}}
\def\M{{\mathbb M}}         \def\ofi{\overline{\phi}}
         \def\oI{{\overline I}}
\def\R{{\mathbb R}}         \def\oJ{{\overline J}}
\def\S{{\mathbb S}}         \def\olam{\overline{\lambda}}
\def\T{{\mathbb T}}         
\def\V{{\mathbb V}}         \def\ophi{\overline{\varphi}}
         
                            \def\uc{{\underline c}}
\def\cA{{\cal A}}           \def\vY{\vec Y{}}
           \def\vfi{\!\vec{\,\varphi}}

\def\cR{{\cal R}}           \def\Qsq{Q_s^2}
\def\cS{{\cal S}}           \def\Ysq{Y_s^2}

\def\Lsc{L_{\rm sc}}                \def\ME {\mbox{$\M_{\rm E}$}}
           \def\MED {\mbox{$\M_{\rm E}^D$}}
           \def\MJ {\mbox{$\M_{\rm J}$}}
\def\Mext {\mbox{$\M_{\rm ext}$}}   \def\MJD {\mbox{$\M_{\rm J}^D$}}
\def\umx{u_{\max}}
\def\Str{\mbox{$\S_{\rm trans}$}}       \def\m{{\rm m}}
                                        
                   \def\Fei{F_{\e I}}
               \def\Fmi{F_{\m I}}
\def\sums{\sum_s}                       
\def\sumt{\sum_{i=2}^{n}}               

\def\bopr{{\beta^1}'}
\def\boprr{{\beta^1}''}

\def\Df{\Delta\phi}
\def\rank{\mathop{\rm rank}\nolimits}

\def\GR{general relativity}                \def\Tan{Tangherlini}
\def\sph{spherically symmetric}            \def\dS{de Sitter}
\def\ssph{static, spherically symmetric}   \def\AdS{anti-de Sitter}
                           \def\Sch{Schwa\-rz\-schild}
\def\bh{black hole}                        \def\brane{$p$-brane}
\def\bhs{black holes}                      \def\branes{$p$-branes}
                          \def\bw{brane world}
\def\whs{wormholes}                        \def\bwd{brane-world}
\def\asflat{asymptotically flat}           \def\TH{T_{{}_{\rm H}}}

\heads  {K.A. Bronnikov and V.N. Melnikov}
{Conformal frames and $D$-dimensional gravity }

\begin{document}
\thispagestyle{empty}
\rightline{\bf gr-qc/0310112}
\bigskip

\Title {CONFORMAL FRAMES AND D-DIMENSIONAL GRAVITY}

\Author{K.A. Bronnikov and V.N. Melnikov}
{Centre for Gravitation and Fundamental Metrology, VNIIMS,\\
      \cm 3-1 M. Ulyanovoy St., Moscow 119313, Russia;\\
Institute of Gravitation and Cosmology, PFUR,
             6 Miklukho-Maklaya St., Moscow 117198, Russia}

\Abstract
     {We review some results concerning the properties of \ssph\
     solutions of multidimensional theories of gravity: various
     scalar-tensor theories and a generalized string-motivated model
     with multiple scalar fields and fields of antisymmetric forms
     associated with \branes. A Kaluza-Klein type framework is used:
     there is no dependence on internal coordinates but multiple
     internal factor spaces are admitted.
     We discuss the causal structure and the existence of black holes,
     wormholes and particle-like configurations in the case of
     scalar vacuum with arbitrary potentials as well as some observational
     predictions for exactly solvable systems with \branes:
     post-Newtonian coefficients, Coulomb law violation and \bh\
     temperatures. Particular attention is paid to conformal frames
     in which the theory is initially formulated and which are used
     for its comparison with observations; it is stressed that,
     in general, these two kinds of frames do not coincide.
}

\section{Introduction}

     The known gravitational phenomena are rather well described in the
     framework of conventional \GR\ (GR). However, in a more
     general context of theoretical physics, whose basic aims are to
     construct a ``theory of everything'' and to explain why our Universe
     looks as it looks and not otherwise, most of the recent advances are
     connected with models in dimensions greater than four: Kaluza-Klein
     type theories, 10-dimensional superstring theories, M-theory and their
     further generalizations. Even if such theories (or some of them)
     successfully explain the whole wealth of particle and astrophysical
     phenomenology, there remains a fundamental question of finding direct
     observational evidence of extra dimensions, which is of utmost
     importance for the whole human world outlook.

     Observational ``windows'' to extra dimensions are discussed
     for many years \cite{M0}--\cite{M2}. Among the well-known predictions
     are variations of the fundamental physical constants on the
     cosmological time scale \cite{will}--\cite{mel6}. Such constants are,
     e.g., the effective gravitational constant $G$ and the fine structure
     constant $\alpha$.  There exist certain observational data on $G$
     stability on the level of $\Delta G/G \sim 10^{-11}\div 10^{-12}$
     y$^{-1}$ \cite{will,mel3,damr}, which restrict the range of viable
     cosmological models.  Some evidence on the variability of $\alpha$ has
     also appeared from quasar absorption spectra: $\Delta \alpha/\alpha
     \sim-0.72\ten{-5}$ over the redshift range $0.5 < z  < 3.5$
     \cite{alpha-} (the minus means that $\alpha$ was smaller in the past).

     Other possible manifestations of extra dimensions include excitations
     in compactified factor spaces \cite{zhuk}, which can behave as
     particles with a large variety of masses and contribute to dark matter
     or to cross-sections of usual particle interactions; monopole modes in
     gravitational waves; various predictions for standard cosmological
     tests and generation of the cosmological constant \cite{GIM}, and
     numerous effects connected with local field sources, including, in
     particular, deviations from the Newton and Coulomb laws
     \cite{mel4,mel5,M1,bm-ann,M3I} and the properties of \bhs, especially in
     the actively discussed \bwd\ framework \cite{BW}.

     In this paper we discuss solutions of multidimensional theories of
     gravity of Kaluza-Klein type, i.e., under the condition that neither
     the metric nor other fields depend on the additional (internal)
     coordinates \cite{M0}--\cite{M2}. The 4D metric is generally specified
     in such theories up to multiplying by a conformal factor depending on
     scalar fields and extra-dimension scale factors. This is the well-known
     problem of choice of a physical conformal frame (CF). Mathematically, a
     transition from one CF to another is nothing else but a substitution in
     the field equations, which can be solved using any variables. However,
     physical predictions about the behaviour of matter (except massless
     particles) are CF-dependent.

     Among possible CFs one is distinguished: the so-called Einstein frame,
     in which the metric field Lagrangian contains the scalar
     curvature $\cR$ with a constant coefficient. In other, so-called Jordan
     frames, $\cR$ appears with field-dependent factors.

     The choice of a physical CF in non-Einsteinian theories of gravity is
     rather widely discussed, but mostly in four dimensions in the context of
     scalar-tensor theories (STT) and in higher-order theories with
     curvature-nonlinear gravitational Lagrangians --- see, e.g.,
     \cite{sokol93,fara98,rzhuk98} and numerous references therein. The
     review \cite{fara98} classified the authors of published papers by their
     attitude to the problem: those (i) neglecting the issue; (ii) supporting
     the view that all frames are equivalent; (iii) recognizing the problem
     but giving no conclusive arguments; (iv) claiming that a Jordan frame
     is physical; (v) asserting that the Einstein frame is physical. Each
     group included tens of names, and some names even got into more than
     one group.

     Refs. \cite{sokol93} and \cite{fara98} have presented arguments in
     favour of the Einstein frame, and the most important ones, applicable
     to STT and higher-order theories (and multiscalar-tensor theories
     obtainable from multidimensional gravity) are connected with the
     positivity of scalar field energy and the existence of a
     classically stable ground state.

     In our view \cite{mel3,br95-2,pn} [which turns out to be outside the
     groups (i)--(v)], the above arguments could be convincing if we
     dealt with an ``absolute'', or ``ultimate'' theory of gravity. If,
     however, the gravitational action is obtained in a certain limit of a
     more fundamental unified theory, theoretical requirements like the
     existence of a stable ground state should be addressed to this
     underlying theory rather than its visible manifestation. In the latter,
     the notion of a physical CF should be only related to the properties of
     instruments used for measuring masses, lengths and time intervals.
     Moreover, different sets of instruments (different measurement systems
     \cite{mel3}) are described, in general, by different CFs. We thus
     suppose that there can be at least two different physical CFs:
     the fundamental one, in which the underlying field theory (or a
     field limit of a more fundamental theory) is specified,
     and the observational one, corresponding to a given set of instruments.
     One can say that the first CF describes what is happening
     ``as a matter of fact'', the second one --- what we see.

     The set of references used in the present observations is connected
     with atomic units, and the corresponding observational CF for any
     underlying theory is therefore the CF that provides geodesic motion
     for ordinary massive (fermionic) matter in 4 dimensions.

     In what follows, we will first discuss the CF dependence of the
     properties of space-time using, as an example, \ssph\ scalar field
     configurations in STT (\sect 2) and in multidimensional theories of
     gravity with multiple factor spaces (\sect 3). We shall see that
     some general theorems, valid in one CF, may be violated in another, and
     there are such conformal mappings that the space-times of different
     frames are even not in a one-to-one correspondence (the so-called
     conformal continuation \cite{vac4}). Then, in \sect 4, we will discuss
     the CF dependence of some observable quantities for a class of
     solutions of a generalized field model \cite{IM0}--\cite{bim97},
     containing multiple scalar fields and antisymmetric forms, associated
     with charged \branes. This choice is motivated by the bosonic sector of
     the low-energy field approximation of superstring theories, M-theory
     and their generalizations \cite{GSW}--\cite{khv}. The model is,
     however, not restricted to known theories since it assumes arbitrary
     dimensions of factor spaces, arbitrary ranks of antisymmetric forms and
     an arbitrary number of scalar fields. Among the quantities to be
     discussed are (1) the post-Newtonian (PN) coefficients describing the
     weak field behaviour of the solutions, (2) for \bh\ solutions, the
     Hawking temperature $\TH$ which is obviously important for
     small (e.g., primordial) \bhs\ rather than those of the stellar or
     galactic mass range and (3) the parameters of Coulomb law violation for
     the 4D components of the antisymmetric forms which behave as an
     electromagnetic field. We do not fix the underlying fundamental theory
     and thus have no reason to prescribe a particular CF, therefore the
     results are formulated in an arbitrary frame.

\section {Scalar-vacuum configurations in STT}

\subsection{STT in Jordan and Einstein frames}

     Consider a general (Bergmann-Wagoner-Nordtvedt) STT, in a
     $D$-dimen\-si\-onal manifold $\MJ[g]$ with the metric $g\mn$
     (to be called the Jordan conformal frame),
     for which the gravitational field action is written of the form
\bearr
     S_{\rm STT} = \int d^D x \sqrt{g}                    \label{act-J}
                  [f(\phi) \cR + h(\phi) (\d\phi)^2 -2U(\phi)],
\ear
     where $g =|\det(g\mn)|$,
     $(d\phi)^2 = g\MN\,(\d_{\mu}\phi)(\d_{\nu}\phi)$ and
     $f,\ h,\ U$ are arbitrary functions of the scalar field $\phi$.

     The action (\ref{act-J}) can be simplified by the well-known
     conformal mapping which generalizes Wagoner's \cite{wagon} 4-dimensional
     transformation,
\bear
     g\mn \eql F(\phi) \og\mn, \cm
                 F(\phi) := |f(\phi)|^{-2/(D-2)},           \label{g-wag}
\\
     \frac{d\psi}{d\phi} \eql \pm
                \frac{\sqrt{|l(\phi)}|}{f(\phi)},         \label{phi-wag}
\cm
     l(\phi) := fh + \frac{D-1}{D-2}\Bigl(\frac{df}{d\phi}\Bigr)^2,
\ear
     removing the nonminimal scalar-tensor coupling express\-ed in the
     factor $f(\phi)$ before $\cR$. The action (\ref{act-J}) is now
     specified in the new manifold $\ME[\og]$ with the metric $\og\mn$
     (the Einstein frame) and the new scalar field $\psi$:
\bearr
     S_{\rm E} = \int  d^D x \sqrt{\og}                        \label{act-E}
        \Bigl\{\sign f \bigl[\ocR
                + (\sign l) (\d\psi)^2\bigr] -2V(\psi)\Bigr\},
\ear
     where the determinant $\og$, the scalar curvature $\ocR$ and
     $(\d\psi)^2$ are calculated using $\og\mn$, and
\beq
        V(\psi) = |f(\phi)|^{-D/(D-2)}\, U(\phi).             \label{UV}
\eeq
     The action (\ref{act-E}) is similar to that of GR with a minimally
     coupled scalar field $\psi$ but, in addition to arbitrary $D$, contains
     two sign factors. The usual sign of gravitational coupling corresponds
     to $f > 0$. On the other hand, theories with $l(\phi) < 0$ lead
     to an anomalous sign of the kinetic term of the $\psi$ field in
     (\ref{act-E}) --- a ``ghost'' scalar field as it is sometimes called.
     Such fields violate all standard energy conditions and therefore easily
     lead to unusual solutions like wormholes \cite{br73,h_ellis}. We will
     adhere to theories with $l > 0$. However, $f < 0$ in some regions
     of \MJ\ will appear due to continuations to be discussed further.

     Among the three functions of $\phi$ entering into (\ref{act-J})
     only two are independent since there is a freedom of transformations
     $\phi = \phi(\phi_{\rm new})$. We assume $h\geq 0$ and use this
     freedom, choosing in what follows $h(\phi) \equiv 1$.

\subsection{No-go theorems for the Einstein frame}

     Let us discuss some general properties of \ssph\ scalar-vacuum
     configurations. We begin with the Einstein frame $\ME(\og)$ with
     the action (\ref{act-E}) and put $\sign f = \sign l = 1$, thus
     obtaining $D$-dimensional GR with a minimally coupled scalar field
     $\psi$.

     Choosing the radial coordinate $\rho$ corresponding to the gauge
     condition $\og_{tt}\,\og_{\rho\rho} = -1$, we can write an arbitrary
     \ssph\ metric in the form
\beq                                                           \label{dsE}
     ds_{\rm E}^2 = A(\rho)\, dt^2 - \frac{du^2}{A(\rho)}
                                       - r^2(\rho)\, d\Omega_{d_0}^2
\eeq
     where $d_0 = D-2$ and $d\Omega_{d_0}^2$ is the linear element on the
     sphere $\S^{d_0}$ of unit radius. This gauge is preferable for
     considering Killing horizons, described as zeros of the function
     $A(\rho)$. The reason is that near a horizon $\rho$ varies (up to a
     positive constant factor) like manifestly well-behaved Kruskal-like
     coordinates used for an analytic continuation of the metric. Thus,
     using this coordinate, which may be called {\it quasiglobal\/}
     \cite{vac1}, one can ``cross the horizons'' preserving the formally
     static expression for the metric.

     Three independent field equations due to (\ref{act-E}) for the unknowns
     $A(\rho)$, $r(\rho)$ and $\psi(\rho)$ may be written as follows:
\bear
              (A'r^{d_0})' \eql - (4/d_0)r^{d_0} V;               \label{00d}
\\
              d_0 r''/r  \eql -{\psi'}^2;                     \label{01d}
\\
      A (r^2)'' - r^2 A''\eql (d_0-2) r' (A'r - 2Ar')
                     + 2(d_0-1),              \label{02d}
\ear
     where the prime denotes $d/d\rho$. These are three combinations of the
     Einstein equations; the scalar field equation  $(Ar^{d_0}\psi')' =
     r^{d_0} dV/d\psi$ can be obtained as their consequence.

     \eqs (\ref{00d})--(\ref{02d}) cannot be exactly solved for a given
     arbitrary potential $V(\psi)$ but make it possible to prove some
     important theorems telling us what can and what cannot be expected from
     such a system:

\begin{description}    \itemsep 1pt
\item[A.] The no-hair theorem \cite{bek,ad-pear} claiming that
     \asflat\ \bhs\ cannot have nontrivial external scalar fields with
     nonnegative $V(\psi)$. In other words, in case $V\geq 0$, the only
     \asflat\ \bh\ solution is characterized outside
     the horizon by $V\equiv 0$, $\psi = \const$ and the \Sch\ (or
     Tangherlini in case $d_0 > 2$) metric, i.e., in (\ref{dsE}) $r \equiv
     \rho$ and $A = A(r) = 1 - 2m/r^{d_0-1}$, $m=\const$.

\item[B.] The generalized Rosen theorem \cite{brsh91,vac5} asserting that
     particle-like solutions (i.e., \asflat\ solutions with a regular
     centre) do not exist in case $V\geq 0$.

\item[C.] The nonexistence theorem for regular configurations without a
     centre (wormholes, horns, flux tubes with $\psi \ne \const$)
     \cite{vac1}.

\item[D.] The causal structure theorem \cite{vac1}, asserting that
     the list of possible types of global causal structures
     (described by Carter-Penrose diagrams) for configurations with any
     potentials $V(\psi)$ and any spatial asymptotics is the same as the
     one for $\psi = \const$, namely: Minkowski (or AdS), \Sch,
     de Sitter and \Sch{}--de Sitter.
\end{description}

    These results will be referred to as Statements A, B, C, D, respectively.

    Some comments are in order. Statement A is proved \cite{bek,ad-pear}
    (see also \cite{vac5} for $D > 4$) by finding an integral relation
    whose two parts have different signs unless the scalar field is trivial.
    There also exist no-hair theorems for \bhs\ with \dS\ and \AdS\
    asymptotics \cite{NHdS}.

    Statement B is proved in its most general form \cite{vac5} by comparing
    two expressions for the mass: one written as an integral of the energy
    density and another given by the Tolman formula.

    In Statement C, a {\it wormhole\/} is, by definition, a configuration
    with two asymptotics at which $r(\rho)\to \infty$, hence $r(\rho)$ must
    have at least one regular minimum. A {\it flux tube\/} is characterized
    by $r = \const >0$, i.e., it is a static $(d_0+1)$-dimensional cylinder.
    A {\it horn\/} is a configuration that tends to a flux tube at one of
    its asymptotics, i.e., $r(\rho)\to\const >0$ at one of the ends of the
    range of $\rho$. The statement is proved \cite{vac1} using \eq
    (\ref{01d}):  e.g., it leads to $r'' \leq 0$, which is incompatible with
    a regular minimum of $r(\rho)$.

    A proof of Statement D \cite{vac1,vac2} rests on \eq (\ref{02d}) which
    implies that the function $A(\rho)/r^2$ cannot have a regular minimum,
    therefore $A(\rho)$ can have at most two simple zeros around a static
    (R) region with $A>0$ or one double zero separating two nonstatic (T)
    regions ($A<0$).

    It should be stressed that the validity of Statements C and D is
    independent of any assumptions on the shape and even sign of the
    potential $V(\psi)$ and on the particular form of the spatial asymptotic.

    In cases admitted by the above theorems, \bh\ and particlelike solutions
    can be obtained, as is confirmed by known explicit examples. Thus, there
    exist: (1) \bhs\ possessing nontrivial scalar fields (scalar hair), with
    $V\geq 0$, but with non-flat and non-de Sitter asymptotics
    \cite{Mann95}; (2) \bhs\ with scalar hair and flat asymptotics, but
    partly negative potentials \cite{vac2}; (3) configurations with a
    regular centre, a flat asymptotic and positive mass, but also with
    partly negative potentials \cite{vac2}.

\subsection{No-go theorems for generic scalar-tensor solutions}

    In this section we discuss the possible validity of Statements A--D for
    STT solutions in a Jordan frame.

    One can notice that when a space-time manifold $\ME[\og]$ (the Einstein
    frame) with the metric (\ref{dsE}) is conformally mapped into another
    manifold $\MJ[g]$ (the Jordan frame) equipped with the same coordinates
    according to the law (\ref{g-wag}), then a horizon $\rho=h$ in \ME\
    passes into a horizon of the same order in \MJ, a centre ($r=0$) and an
    asymptotic ($r\to \infty$) in \ME\ pass into a centre and an asymptotic,
    respectively, in \MJ\ if the conformal factor $F=F(\rho)$ is regular
    (i.e., finite, at least C${}^2$-smooth and positive) at the
    corresponding values of $\rho$. A regular centre passes to a regular
    centre and a flat asymptotic to a flat asymptotic under evident
    additional requirements.

    The validity of Statements A--D in the Jordan frame thus depends on the
    nature of the conformal mapping (\ref{g-wag}) that connects $\MJ[g]$
    with $\ME [\og]$). Thus, if $F$ vanishes or blows up at an intermediate
    value of $\rho$, there is no one-to-one correspondence between \MJ\ and
    \ME. In particular, if a singularity in \ME\ is mapped to a regular
    sphere \Str\ in \MJ, then \MJ\ should be continued beyond this sphere,
    and we obtain, by definition, a {\it conformal continuation\/} (CC) from
    \ME\ into \MJ\ \cite{vac3,vac4}.

    Such continuations can only occur for special solutions: to be removed
    by a conformal factor, the singularity should be, in a sense, isotropic.
    Moreover, the factor $F$ should behave precisely as is needed to
    remove it.

    In more generic situations, for given \ME, there is either a one-to-one
    correspondence between the two manifolds, or the factor $F$ ``spoils''
    the geometry and creates a singularity in \MJ, that is, in a sense,
    \MJ\ is ``smaller'' than \ME. In these cases Statement
    D is obviously valid in \MJ. This is manifestly true for STT with
    $f(\phi) >0$.

    Statement C cannot be directly transferred to \MJ\ in any nontrivial
    case $F \ne \const$. In particular, minima of $g_{\theta\theta}$
    (wormhole throats) can appear. Though, wormholes as global entities are
    impossible in \MJ\ if the conformal factor $F$ is finite in the whole
    range of $\rho$, including the boundary values. Indeed, assuming
    that there is such a wormhole, we shall immediately obtain
    two large $r$ asymptotics and a minimum of $r(\rho)$ between them even
    in \ME, in contrast to Statement C valid there.

    Statements A and B can also be extended to \MJ\ for generic
    STT solutions, but here we will not concentrate on the details
    and refer to the papers \cite {NH-STT} (see also \sect 3).

    Conformal continuations, if any, can in principle lead to new, maybe
    more complex structures.

\subsection{Conformal continuations}

    A CC from \ME\ into \MJ\ can occur at such values of the scalar field
    $\phi$ that the conformal factor $F$ in the mapping (\ref{g-wag}) is
    singular while the functions $f$, $h$ and $U$ in the action
    (\ref{act-J}) are regular.
    This means that at $\phi=\phi_0$, corresponding to a
    possible transition surface \Str, the function $f(\phi)$ has a zero of a
    certain order $n$. Then, in the transformation (\ref{phi-wag})
    near $\phi=\phi_0$ in the leading order of magnitude
\beq
      f(\phi) \sim \Df^n, \qquad   n = 1,2,\ldots,
         \cm        \Df \equiv  \phi-\phi_0.      \label{phi0}
\eeq

    One can notice, however, that $n > 1$ leads to $l(\phi_0) = 0$
    (recall that by our convention $h(\phi)\equiv 1$). This generically
    leads to a curvature singularity in \MJ, as can be seen from the
    trace of the metric field equation due to (\ref{act-J}) \cite{vac4}.
    We therefore assume $l > 0$ at \Str. Therefore, according to
    (\ref{phi-wag}), we have near \Str\ ($\phi=\phi_0$):
\bear
    f (\phi) \sim \Df \sim \e^{-\psi\sqrt{d_0/(d_0+1)}},       \label{Df}
\ear
    where without loss of generality we choose the sign of $\psi$ so that
    $\psi\to\infty$ as $\Df \to 0$.

    In the CC case, the metric $\og\mn$ is singular on \Str\ while $g\mn =
    F(\phi)\,\og\mn$ is regular. There are two opportunities. The first one,
    to be called CC-I for short, is that \Str\ is an {\it ordinary regular
    surface\/} in \MJ, where both $g_{tt}= \cA = FA$ and $-g_{\theta\theta}
    = R^2 = Fr^2$ (squared radius of \Str) are finite.  (Here $\theta$ is
    one of the angles that parametrize the sphere $\S^{d_0}$.) The second
    variant, to be called CC-II, is that \Str\ is a {\it horizon} in \MJ. In
    the latter case only $g_{\theta\theta}$ is finite, while $g_{tt}=0$.

    Given a metric $\og\mn$ of the form (\ref{dsE}) in \ME, a CC-I
    can occur if
\beq                                                          \label{CC-g}
    F(\psi) = |f|^{-2/d_0} \sim 1/r^2 \sim 1/A
\eeq
    as $\psi\to\infty$, while the behaviour of $f$ is specified by
    (\ref{Df}). In \ME, the surface \Str\ ($r^2 \sim A \to 0$) is either
    a singular centre, if the continuation occurs in an R-region, or
    a cosmological singularity in the case of a T-region.

    It has been shown \cite{vac4} that necessary and sufficient conditions
    for the existence of CC-I are that $f(\phi)$ has a simple zero at some
    $\phi = \phi_0$, and $|U(\phi_0)| < \infty$.
    Then there is a solution in \MJ, smooth in a neighbourhood of the
    surface \Str\ ($\phi=\phi_0$), and in this
    solution the ranges of $\phi$ are different on different sides of \Str.

    Thus any STT with $h\equiv 1$ admitting a simple zero of
    $f(\phi)$ admits a CC-I. The smooth solution in \MJ\ corresponds to
    two solutions on different sides of \Str\ in two different Einstein
    frames. These solutions are special, being restricted by \eq
    (\ref{CC-g}).

    It is of interest that, under the CC-I conditions, any finite potential
    $V(\psi)$ is inessential near \Str: the solution is close to Fisher's
    scalar-vacuum solution \cite{fisher} for $D=4$ or its modification
    in other dimensions. For $U(\phi)$, the CC-I conditions do not
    lead to other restrictions than regularity at $\phi=\phi_0$.

    In case $D=3$, as follows from \eq (\ref{02d}), a necessary condition
    for CC-I is $A/r^2 = \const$.

    A CC-II requires more special conditions \cite{vac4}, namely, there
    should be $D\geq 4$, $U(\phi) = 0$ and $dU/d\phi \ne 0$ at $\phi=\phi_0$.
    It then follows that \Str\ is a second-order horizon, connecting two
    T-regions in \MJ. Thus the only kind of STT configurations admitting
    CC-II is a $D\geq 4$ Kantowski-Sachs cosmology consisting of two
    T-regions (in fact, epochs, since $\rho$ is then a temporal coordinate),
    separated by a double horizon.

\subsection{Global properties of continued solutions}

    A solution to the STT equations may {\it a priori\/} undergo a number of
    CCs, so that each region of \MJ\ between adjacent surfaces \Str\ is
    conformally equivalent to some \ME. However, the global properties of
    \MJ\ with CCs are not so diverse as one might expect. In particular,
    Statement D, restricting possible causal structures, holds in \MJ\
    in the same form as in \ME.

    A key point for proving this is the observation that the quantity
    $B = A/r^2$ is insensitive to conformal mappings (both its numerator and
    denominator are multiplied by $F$) and is thus common to \MJ\ and \ME\
    which is equivalent to a given part of \MJ. Therefore zeros and extrema
    of $B$ inside \ME\ preserve their meaning in the corresponding part of
    \MJ. Statement D rests on the fact that $B(\rho)$ cannot have a regular
    minimum in \ME; the same is true in a region of \MJ\
    equivalent to some \ME, and a minimum can only take place on a transition
    surface \Str\ between such regions. A direct inspection shows
    \cite{vac4} that this is not the case. Therefore Statement D is valid in
    \MJ\ despite any number of CC's.

    As for Statements A--C, the situation is more involved. To our
    knowledge, full analogues of Statements A and B (probably with
    additional restrictions) for a sufficiently general STT are yet
    to be obtained (see, however, \cite{NH-STT}).
    Statement C is evidently violated due to CCs since \whs\ are
    a generic product of such continuations \cite{vac4}.

    Indeed, a generic behaviour of \ME\ is that $r$ varies from zero to
    infinity. Let there be a family of such static solutions and let
    $f(\phi)$ have a simple zero. Then there is a subfamily of solutions
    admitting CC-I. A particular solution from this subfamily can come
    across a singularity beyond \Str\ [due to $f(\phi) \to \infty$ or
    $l(\phi) \to 0$], but if ``everything is quiet'', it will, in general,
    arrive at another spatial asymptotic and will then describe a wormhole.

    It can be shown \cite{vac4} that, under our assumption $l > 0$, there
    cannot be more than two values of $\phi$ where CCs are possible, i.e.,
    where $f=0$ and $df/d\phi\ne 0$. This does not mean, however, that an
    STT solution cannot contain more than two CCs. The point is that $\phi$
    as a function of the radial coordinate is not necessarily monotonic, so
    there can be two or more CCs corresponding to the same value of $\phi$.
    A transition surface $\Str \in \MJ$ corresponds to $r=0$
    in \ME, therefore an Einstein-frame manifold \ME, describing a region
    between two transitions, should contain two centres, more precisely,
    two values of the radial coordinate (say, $\rho$) at which $r=0$.
    This property, resembling that of a closed cosmological model, is quite
    generic due to $r''\leq 0$ in \eq (\ref{01d}), but a special feature is
    that the conditions (\ref{CC-g}) should hold at both centres.

    Well-known particular examples of CC-I are connected with
    massless nonminimally coupled scalar fields in GR, which may
    be described as STT with $f(\phi) = 1-\xi \phi^2$, $h(\phi) = 1$,
    $U(\phi) = 0$. One such example is a \bh\ with a conformally
    coupled field ($\xi = 1/6$)) \cite{bbm70,bek}, such that $\phi=\infty$
    but the energy-momentum tensor is finite on the horizon. Other examples
    are \whs\ supported by conformal \cite{br73} and nonconformal
    \cite{viss} fields. Ref.\,\cite{vac4} contains an example of a
    configuration with an infinite number of CCs, built using
    a conformally coupled scalar field with a nonzero potential in
    three dimensions.

\section{Theories with multiple factor spaces}

\subsection{Reduction}

     In \sect 2 we have been concerned with STT solutions in $D$-dimensional
     space-times with the metrics $\og\mn$ given by (\ref{dsE}) and
     $g\mn = F(\phi)\,\og\mn$. Let us now pass to space-times $\M^D$ with a
     more general structure
\beq                                                         \label{M-prod}
     \M^D = \R_u \times\M_0\times\M_1\times \M_2 \times \cdots \times \M_n
\eeq
     where $\Mext = \R_u \times\M_0 \times\M_1$ is the ``external''
     manifold, $\R_u \subseteq \R$ is the range of the radial coordinate
     $u$, $\M_1$ is the time axis, $\M_0 = \S^{d_0}$. Furthermore, $\M_2,
     \ldots, \M_n$ are ``internal'' factor spaces of arbitrary dimensions
     $d_i$, $i=2,\ldots, n$, and, according to this notation, we also have
     $\dim \M_0 = d_0$ and $\dim \M_1 = d_1 =1$. The metric is taken in
     the form
\bearr                                                         \label{dsDn}
     ds_D^2 = - \e^{2\alpha^0} du^2 - \e^{2\beta^0} d\Omega_{d_0}^2
              + \e^{2\beta^1} dt^2 - \sum_{i=2}^{n}\e^{2\beta^i} ds_i^2,
\ear
     where $ds_i^2$ ($i=2, \ldots, n$) are metrics of Einstein spaces of
     arbitrary dimensions $d_i$ and signatures while $\alpha^0$ and all
     $\beta^i$ are functions of the radial coordinate $u$.

     Consider in $\M^D$ a field theory with the action
\beq
     S = \int d^D x \,\sqrt{|g_D|}                            \label{act-D}
                    \Bigl[\cR_D
     h_{ab}(\ofi) g^{MN}(\d_M\phi^a)(\d_N \phi^b) - 2 V_D(\ofi)\Bigr],
\eeq
     where $\cR_D$ is the $D$-dimensional scalar curvature and
     the scalar field Lagrangian has a $\sigma$-model form.
     We assume that $\phi^a$ are functions of the external space coordinates
     $x^{\mu}$ ($\mu = 0, 1, ..., d_0+1$), so that actually in (\ref{act-D})
     $g^{MN}(\d_M\phi^a)(\d_N \phi^b) = g\MN (\d_\mu \phi^a) (\d_\nu
     \phi^b) \equiv (\d\phi^a, \d\phi^b)$, where the metric $g\mn$ is formed
     by the first three terms in (\ref{dsDn}). The metric $h_{ab}$ of the
     $N'$-dimensional target space $\T_{\phi}$, parametrized by $\phi^a$,
     and the potential $V$ are functions of $\ofi = \{\phi^a\} \in
     \T_{\phi}$.

     The action (\ref{act-D}) represents in a general form the scalar-vacuum
     sector of diverse supergravities and low-energy limits of string
     and $p$-brane theories \cite{GSW,brane}. In many papers devoted to
     exact solutions of such low-energy theories (see \sect 4) all internal
     factor spaces are assumed to be Ricci-flat, and nonzero potentials
     $V_D(\ofi)$ are not introduced due to technical difficulties of solving
     the equations.  Meanwhile, the inclusion of a potential not only makes
     it possible to treat massive and/or nonlinear and interacting scalar
     fields, but is also necessary for describing, e.g., the symmetry
     breaking and Casimir effects. (On the use of effective potentials for
     describing the Casimir effect in compact extra dimensions, see, e.g.,
     \cite{zhuk} and references therein.)

     Let us perform a dimensional reduction to the external space-time
     \Mext\ with the metric $g\mn$. \eq (\ref{act-D}) is converted to
\bearr                                                     \label{act-Dex}
        S = \int d^{d_0+2}x \sqrt{|g_{d_0+2}|} \e^{\sigma_2}
            \biggl\{   \cR_{d_0+2}
                 + \sum_{i=2}^{n} d_i(d_i-1)K_i\e^{-2\beta^i}
\nnn \cm
            + 2 \nabla^\mu \nabla_\mu \sigma_2
     + \sum_{i,k =2}^{n}
            (d_i d_k + d_i\delta_{ik})(\d\beta^i,\d\beta^k) + \Lsc \biggr\},
\ear
     where all quantities, including the scalar $\cR_{d_0+2}$, are
     calculated with the aid of $g\mn$, and
            $\sigma_2:= \sum_{d=2}^{n}d_i\beta^i$,
     so that $\e^{\sigma_2}$ is the volume factor of extra dimensions.

     It is helpful to pass in the action (\ref{act-J}), just as in the
     STT (\ref{act-J}), from the Jordan-frame metric $g\mn$ in \Mext\ to
     the Einstein-frame metric
\beq
     \og\mn = \e^{2\sigma_2/d_0}g\mn.                       \label{conf}
\eeq
     Then, omitting a total divergence, one obtains the
     action (\ref{act-D}) in terms of $\og\mn$:
\bearr \nhq
     S = \int d^{d_0+2} x \,\sqrt{|\og|}                    \label{act-Es}
            \Bigl[\ocR + H_{KL} (\d\varphi^K,\d\varphi^L)
                        -2 V (\vfi)\Bigr].
\ear
     Here the set of fields $\{\varphi^K\} = \{\beta^i,\,\phi^a\}$,
     combining the scalar fields from (\ref{act-D}) and the moduli fields
     $\beta^i$, is treated as a vector in the extended $N =
     (n{-}1{+}N')$-dimensional target space $\T_{\varphi}$ with the metric
\beq
     (H_{KL}) = \pmatrix {d_i d_k/d_0 + d_i\delta_{ik}  &   0\cr
                  0                      & h_{ab}\cr
             },                                               \label{H-KL}
\eeq
     while the potential $V(\varphi)$ is expressed in terms of $V_D(\ofi)$
     and $\beta^i$:
\beq    \nq                                                     \label{V}
     V(\vfi) = \e^{-2\sigma_2/d_0}\biggl[ V_D(\ofi)
              - \Half \sum_{i=2}^{n} K_i d_i(d_i{-}1)\e^{-2\beta^i}\biggr].
\eeq

\subsection{Extended no-go theorems}

     The action (\ref{act-Es}) brings the theory (\ref{act-D}) to a form
     quite similar to (\ref{act-E}) ($f>0$), but a single field $\psi$ is
     now replaced by a $\sigma$ model with the target space metric $H_{KL}$.
     It can be easily shown \cite{vac5} that Statements A--D are entirly
     extended to the theory (\ref{act-Es}) under the condition that the
     metric $H_{KL}$ is positive-definite, which is always the case as long
     as $h_{ab}$ is positive-definite.

     Let us now discuss the properties of the $D$-dimensional metric $g_{MN}$
     given by (\ref{dsDn}). Its ``external'' part $g\mn$ is connected with
     $\og\mn$ by the conformal transformation (\ref{conf}). Since the
     action (\ref{act-D}) corresponds to GR in $D$ dimensions, this frame
     may be called the $D$-dimensional Einstein frame, and we will now
     designate the manifold $\M^D$ endowed with the metric $g_{MN}$ as \MED.

     The nonminimal coupling coefficient in the action (\ref{act-J}), being
     connected with the extra-dimension volume factor $\e^{\sigma_2}$, is
     nonnegative by definition, and the solution terminates where
     $\e^{\sigma_2}$ vanishes or blows up. Thus, in contrast to the
     situation in STT, conformal continuations are here impossible: one
     cannot cross a surface, if any, where $\e^{\sigma_2}$ vanishes. Roughly
     speaking, the Jordan-frame manifold $\Mext [g]$ can be smaller but
     cannot be larger than $\Mext [\og]$. If $\sigma_2 \to \pm \infty$ at
     an intermediate value of the radial coordinate, then the transformation
     (\ref{conf}) maps $\Mext[g]$ to only a part of $\Mext[\og]$.

     Asymptotic flatness of the metric $g_{MN}$ in $\MED$
     implies an \asflat\ Einstein-frame metric $\og\mn$ in \Mext\ and finite
     limits of the moduli fields $\beta^i$, $i\geq 2$, at large radii. A
     similar picture is observed with the regular centre conditions: a
     regular centre in $\MED$ is only possible if there is a regular centre
     in $\Mext[\og]$ and $\beta^i$, $i\geq 2$ sufficiently rapidly tend to
     constant values. A horizon in $\MED$ always corresponds to a horizon in
     $\Mext [\og]$. (The opposite assertions are not always true, e.g., a
     regular centre in $\Mext[\og]$ may be ``spoiled'' when passing to
     $g_{MN}$ by an improper behaviour of the moduli fields $\beta^i$.)

     So the global properties of $\Mext[\og]$ and $\Mext[g]$ (and hence
     $\MED$), associated with Statements A--D, are closely related but not
     entirely coincide.

\medskip\noi
     {\bf A.} The {\bf no-hair theorem} can be formulated for $\MED$ as
     follows:

\medskip\noi
{\sl Given the action (\ref{act-D}) with $h_{ab}$ positive-definite and
     a nonnegative potential (\ref{V}) in the space-time $\MED$ with the
     metric (\ref{dsDn}), the only static, \asflat\ \bh\ solution to the
     field equations is characterized in the region of outer communication
     by $\phi^a = \const$, $\beta^i = \const$ ($i = \overline{2,n}$),
     $V(\vfi) \equiv 0$ and the \Tan\ metric $g\mn$.      }

\medskip
     In other words, the only \asflat\ \bh\ solution is given by the \Tan\
     metric in \Mext, constant scalar fields $\phi^a$ and constant moduli
     fields $\beta^i$ outside the event horizon. Note that in this solution
     the metrics $g\mn$ and $\og\mn$ in $\Mext$ are connected by simple
     scaling with a constant conformal factor since $\sigma_2 = \const$.

     Another feature of interest is that it is the potential (\ref{V})
     that vanishes in the \bh\ solution rather than the original potential
     $V_D(\ofi)$ from \eq (\ref{act-D}). Theorem 5 generalizes the
     previously known property of \bhs\ with the metric (\ref{dsDn}) when the
     internal spaces are Ricci-flat and the source is a massless, minimally
     coupled scalar field without a potential \cite{fa91}.

\medskip\noi
     {\bf B. Particle-like solutions:} Statement B is valid in $\MED$
     in the same formulation as previously in \ME, but the condition
     $V\geq 0$ now also applies to the potential (\ref{V})
     rather than $V_D(\ofi)$ from (\ref{act-D}).

\medskip\noi
     {\bf C. Wormholes} and even wormhole throats are impossible with the
     metric $\og\mn$. The conformal factor $\e^{2\sigma_2/d_0}$ in
     (\ref{conf}) removes the prohibition of throats since for $g\mn$ a
     condition like $r''\leq 0$ is no longer valid. However, a wormhole as a
     global entity with two large $r$ asymptotics cannot appear in
     $\MJ=\Mext [g\mn]$ for the same reason as in \sect 2.3.

     Flux-tube solutions with nontrivial scalar and/or moduli fields are
     absent, as before, but horns are not ruled out since the behaviour of
     the metric coefficient $g_{\theta\theta}$ is modified by conformal
     transformations.

     It should be emphasized that all the restrictions mentioned in items
     A-C are invalid if the target space metric $h_{ab}$ is not
     positive-definite.

\medskip\noi
     {\bf D.} The {\bf global causal structure} of any Jordan frame
     cannot be more complex than that of the Einstein frame even in STT,
     where conformal continuations are allowed --- see \sect 2.5.
     The corresponding reasoning of \cite{vac4} entirely applies to \Mext[g]
     and hence to $\MED$.
     The list of possible global structures is again the same as that for
     the \Tan-\dS\ metric. This restriction does not depend
     (i) on the choice and even sign of scalar field potentials, (ii) on the
     nature of asymptotic conditions and (iii) on the algebraic properties of
     the target space metric. It is therefore the most universal property of
     \sph\ configurations with scalar fields in various theories of gravity.

\smallskip
     A theory in $\M^D$ may, however, be initially formulated in another
     conformal frame than in (\ref{act-D}), i.e., with a nonminimal
     coupling factor $f(\ofi)$  before $\cR_D$.
     Let us designate $M^D$ in this case
     as $\MJD$, a $D$-dimensional Jordan-frame manifold. (An example of
     such a construction is the so-called string metric in string theories
     \cite{GSW,brane} where $f$ depends on a dilaton field related to string
     coupling.)  Applying a transformation like (\ref{g-wag}), we can
     recover the Einstein-frame action (\ref{act-D}) in $\MED$, then by
     dimensional reduction pass to $\Mext[g]$ and after one more conformal
     mapping (\ref{conf}) arrive at the $(d_0+2)$ Einstein frame
     $\Mext[\og]$. Addition of the first step in this sequence of reductions
     weakens our conclusions to a certain extent. The main point is that we
     cannot {\it a priori\/} require $f(\ofi) > 0$ in the whole range of
     $\vfi$, therefore conformal continuations (CCs) through surfaces where
     $f=0$ are not excluded.

     Meanwhile, the properties of CCs have only been studied \cite{vac4} for
     a single scalar field in \Mext\ (in the present notation). In our more
     complex case of multiple scalar fields and factor spaces, such a
     continuation through the surface $f(\ofi) =0$ in the multidimensional
     target space $\T_\phi$ can have yet unknown properties.

     One can only say for sure that the no-hair and no-wormhole theorems
     fail if CCs are admitted. This follows from the simplest example of CCs
     in the solutions with a conformal scalar field in GR, leading to \bhs\
     \cite{bbm70,bek74} and wormholes \cite{br73,viss} and known since
     the 70s although the term ``conformal continuation'' was introduced
     only recently \cite{vac3}. A wormhole was shown to be one of the
     generic structures appearing as a result of CCs in STT (\cite{vac4},
     see sect 2.4 of the present paper).

     If we require that the function $f(\ofi)$ should be finite and nonzero
     in the whole range $\R_u$ of the radial coordinate, including its ends,
     then all the above no-go theorems are equally valid in $\MED$ and
     $\MJD$. One should only bear in mind that the transformation
     (\ref{g-wag}) from $\MED$ to $\MJD$ modifies the potential $V_D(\ofi)$
     multiplying it by $f^{-D/(D-2)}$, which in turn affects the explicit
     form of the condition $V\geq 0$, essential for Statements A and B.

     Statement D on possible horizon dispositions and global causal
     structures will be unaffected if we even admit an infinite growth or
     vanishing of $f(\ofi)$ at the extremes of the range $\R_u$. However,
     Statement C will not survive: such a behaviour of $f$ may create a
     wormhole or horn in $\MJD$. A simple example of this kind is a
     ``horned particle'' in the string metric in dilaton gravity of string
     origin, studied by Banks et al. \cite{banks}.

\section{\branes\ and observable effects}

\subsection{The model and the target space $\V$}

     Let us now consider a model which can be associated with \branes\ as
     sources of antisymmetric form fields. Namely, in the space-time
     (\ref{M-prod}) with the metric (\ref{dsDn}), we take,
     as in Refs.\,\cite{IM0}--\cite{bim97}, \cite{bobs,br-jmp}, the model
     action for $D$-dimensional gravity with several scalar dilatonic fields
     $\varphi^a$ and antisymmetric $n_s$-forms $F_s$:
\beq                                                         \label{act-F}
     S = \frac{1}{2\kappa^{2}}
                 \int d^{D}z \sqrt{|g|} \biggl\{
     \cR + \delta_{ab} g^{MN} \d_{M} \varphi^a \d_{N} \varphi^b
                                                    - \sum_{s\in \cS}
     \frac{1}{n_s!} \e^{2 \lambda_{sa} \varphi^a} F_s^2 \biggr\},
\eeq
     where $F_s^2 = F_{s,\ M_1 \ldots M_{n_s}} F_s^{M_1 \ldots M_{n_s}}$;
     $\lambda_{sa}$ are coupling constants; $s \in \cS$, $a\in \cA$, where
     $\cS$ and $\cA$ are some finite sets. Essential differences from
     (\ref{act-D}) are, besides the inclusion of the term with $F_s^2$, that
     (i) the potential $U$ is omitted (i.e., $\varphi^a$ are not
     self-coupled but coupled to $F_s$), (ii) the ``extra'' spaces $\M_i$
     are assumed to be Ricci-flat and (for simplicity) spacelike and
     (iii) the target space metric $h_{ab}$ is Euclidean,
     $h_{ab}=\delta_{ab}$. The ``scale factors" $\e^{\beta^i}$ and the
     scalars $\varphi^a$ are again assumed to depend on $u$ only.

     The $F$-forms (or, more precisely, their particular nonzero components,
     fixed up to permutation of indices, to be labelled with the subscript
     $s$) should also be compatible with spherical symmetry. They
     are naturally classified as
     {\it electric\/} ($\Fei$) and {\it magnetic\/} ($\Fmi$) forms,
     and each of these forms is
     associated with a certain subset $I = \{i_1, \ldots, i_k \}$ ($i_1 <
     \ldots < i_k$) of the set of numbers labelling the factor spaces: $\{i\}
     = I_0 = \{0, \ldots, n \}$. By definition, an
     electric form $\Fei$ carries the coordinate indices $u$ and those
     of the subspaces
     $\M_i,\ i\in I$, whereas a magnetic form $\Fmi$ is built as a
     form dual to a possible electric one associated with $I$.
     Thus nonzero components of $\Fmi$ carry coordinate indices of the
     subspaces $\M_i,\ i\in \oI := I_0 \setminus I$. One can write:
\beq
     n_{\e I} = \rank F_{\e I} = d(I) + 1,\qquad
     n_{\m I} = \rank F_{\m I} = D - n_{\e I} = d(\oI) \label{2.22}
\eeq
     where $d(I) = \sum_{i\in I} d_i = \dim \M_I$,
     $\M_I := \M_{i_1} \times \ldots \times \M_{i_k}$.
     The index $s$ jointly describes the two types of forms.

     If the time axis $\R_t$ belongs to $\M_I$, we are dealing with a true
     electric or magnetic form, directly generalizing the Maxwell field in
     \Mext; otherwise the $F$-form behaves in \Mext\ as an effective scalar
     or pseudoscalar. Such $F$-forms will be called {\sl quasiscalar.\/}

     The forms $F_s$ are associated with \branes\ as extended sources of the
     \sph\ field distributions, where the brane dimension is $p=d(I_s)-1$,
     and $d(I_s)$ is the brane world volume dimension. A natural assumption
     is that the branes only ``live'' in extra dimensions, i.e.,
     $0 \not\in I_s,\ \  \forall s$.

     The classification of $F$-forms can be illustrated using as an example
     $D=11$ supergravity, representing the low-energy limit of
     M-theory \cite{brane}. The action (\ref{act-F}) for the bosonic sector
     of this theory (truncated by omitting the Chern-Simons term) does not
     contain scalar fields, and the only $F$-form is of rank 4, whose
     various nontrivial components $F_s$ (elementary $F$-forms, called
     simply $F$-forms according to the above convention) are associated with
     electric 2-branes [for which $d(I_s)=3$] and magnetic 5-branes [such
     that $d(I_s)=6$, see (\ref{2.22})].

     Let us put $d_0=2$ and ascribe to the external space-time coordinates
     the indices $M = t,\ u,\ \theta,\ \phi$ ($\theta$ and $\phi$ are the
     spherical angles), and let the numbers $i=2,\ldots,8$ refer to the extra
     dimensions, each associated with an extra factor space $\M_i$
     with the same number ($\M_i$ are thus assumed to be one-dimensional).
     The number $i=1$ refers to the time axis, $\M_1 = \R_t$, as stated
     previously. Here are examples of different kinds of forms:
\smallskip

     $F_{ut23}$ is a true electric form, $I = \{123\}$; $\oI = \{045678\}$.

     $F_{\theta\phi 23}$ is a true magnetic form, $I = \{145678\}$;
                     $\oI = \{023\}$.

     $F_{u234}$ is an electric quasiscalar form, $I=\{234\}$;
                     $\oI = \{015678\}$.

     $F_{\theta\phi t 2}$ is a magnetic quasiscalar form, $I=\{345678\}$;
                         $\oI = \{012\}$.
\smallskip

     Under the above assumptions, it is helpful to describe the system in
     the so-called $\sigma$ model representation \cite{IM5}). Namely
     (see more general and detailed descriptions in \cite{M3I,IM5,bim97}),
     let us choose the harmonic
     $u$ coordinate in $\M$ ($\nabla^M \nabla_M u = 0$), such that
\beq                                                         \label{3.1}
     \alpha^0 (u)= \sum_{i=0}^{n} d_i \beta^i
             \equiv d_0\beta^0 + \sigma_1(u).
\eeq
     We use the notations
\beq
     \sigma_i = \sum_{j=i}^{n} d_j\beta^j (u), \cm       \label{sigma}
     \sigma (I) = \sum_{i\in I} d_i\beta^i (u).
\eeq
     Then the combination ${u\choose u}
     {+} {\theta\choose \theta}$ of the Einstein equations, where $\theta$
     is one of the angular coordinates on $\S^{d_0}$, has a Liouville form,
     $\ddot \alpha -\ddot \beta^0 = (d_0{-}1)^2 \e^{2\alpha - 2\beta^0}$
     (an overdot means $d/du$), and is integrated giving
\bear
     \e^{\beta^0 - \alpha^0} = (d_0-1) s(k, u),\cm               \label{3.8}
     s(k,u)  := \vars{      k^{-1} \sinh ku, \quad & k>0,\\
                                     u,        & k=0,\\
                    k^{-1} \sin ku,            & k<0. }
\ear
     where $k$ is an integration constant. Another integration constant is
     suppressed by properly choosing the origin of $u$. With (\ref{3.8}) the
     $D$-dimensional line element may be written in the form
\beq
     ds_D^2 = \frac{\e^{-2\sigma_1/\od}}{[\od s(k,u)]^{2/\od}}
          \Biggl\{ \frac{du^2}{[\od s(k,u)]^2} + d\Omega_{d_0}^2\Biggr\}
              - \e^{2\beta^1} dt^2
                + \sumt \e^{2\beta^i}ds_i^2,              \label{3.10}
\eeq
     $\od := d_0-1$. The range of the $u$ coordinate is  $0 < u < \umx$
     where $u=0$ corresponds to spatial infinity while $\umx$ may be finite
     or infinite depending on the form of a particular solution.

     The Maxwell-like equations for $F_s$ are integrated in a general form,
     giving the respective charges $Q_s = \const$.
     The remaining set of unknowns ${\beta^i(u),\ \varphi^a (u)}$
     ($i = 1, \ldots, n,\ a\in \cA$) can be treated
     as a real-valued vector function $x^A (u)$ (so that
     $\{A\} = \{1,\ldots,n\} \cup \cA$) in an $(n+|\cA|)$-dimensional vector
     space $\V$ (target space). The field equations for $x^A$
     can be derived from the Toda-like Lagrangian
\bearr                                                      \label{3.12}
     L = G_{AB}\dot x^A\dot x^B - V_Q (y), \cm
           V_Q (y) = -\sums \ep_s Q_s^2 \e^{2y_s}
\ear
     (where $\ep_s=1$ for true electric and magnetic form and $\ep_s = -1$
     for quasiscalar forms), with the ``energy'' constraint
\beq                                                      \label{3.16}
     E = G_{AB}\dot x^A \dot x^B + V_Q (y)
                                    = \frac{d_0}{d_0-1}\, k^2 \sign k,
\eeq
     The nondegenerate symmetric matrix
\beq                                                       \label{3.13}
       (G_{AB})=\pmatrix {
           d_id_j/\od + d_i \delta_{ij} &       0      \cr
              0                        &  \delta_{ab} \cr }
\eeq
     specifies a positive-definite metric in $\V$;
     the functions $y_s(u)$ are defined as scalar products:
\beq                                                       \label{3.15}
     y_s = \sigma(I_s) - \chi_s \olam_s\ophi
       \equiv Y_{s,A}  x^A,    \qquad
     (Y_{s,A}) = \Bigl(d_i\delta_{iI_s}, \ \  -\chi_s \lambda_{sa}\Bigr),
\eeq
     where $\delta_{iI} =1$ if $i\in I$ and $\delta_{iI}=0$ otherwise;
     $\chi_s$ distinguish electric and magnetic forms: $\chi_{\e I}=1$,
     $\chi_{\m I} = -1$. The contravariant components and scalar
     products of $\vY_s$ are found using the matrix $G^{AB}$
     inverse to $G_{AB}$:
\bearr                                                      \label{3.18}
     (G^{AB}) = \pmatrix{
            \delta^{ij}/d_i - 1/(D-2) &      0      \cr
      0                               &\delta^{ab}  \cr },
\cm
      (Y_s{}^A) =
       \biggl(\delta_{iI}-\frac{d(I)}{D-2},
                \quad -\chi_s \lambda_{sa}\biggr);
\\
\lal  Y_{s,A}Y_{s'}{}^A \equiv \vY_s \vY_{s'}
                      = d(I_s \cap I_{s'})                \label{3.20}
                      - \frac{d(I_s)d(I_{s'})}{D-2}
                  + \chi_s\chi_{s'} \olam_s \olam_{s'}.
\ear
     The equations of motion in terms of $\vY_s$ read
\beq
     \ddot{x}{}^A = \sums q_s Y_s{}^A \e^{2y_s},
     \cm    q_s  := \ep_s Q_s^2.                       \label{eqm}
\eeq

     One can notice that the metric (\ref{3.13}) is
     quite similar to (\ref{H-KL}), the metric of $\T_{\varphi}$, especially
     if $h_{ab} = \delta_{ab}$. The difference is that in $G_{AB}$ given by
     (\ref{3.13}) we have $i, j = \overline{1,n}$ since $\V$ includes
     as a coordinate the metric function $\beta^1$, whereas in $H_{KL}$
     we have $i,j = \overline{2,n}$, so that $\T_{\varphi}$ is a subspace
     of $\V$.

\subsection{Some exact solutions. Black holes}

     The integrability of the Toda-like system (\ref{3.12}) depends on the
     set of vectors $\vY_s$. Each $\vY_s$ consists of input parameters of
     the problem and represents an $F$-form $F_s$ with a
     nonzero charge $Q_s$, i.e., one of charged \branes.

     The simplest case of integrability takes place
     when $\vY_s$ are mutually orthogonal in $\V$ \cite{bim97}, that is,
\beq                                                        \label{3.21}
     \vY_s \vY_{s'} = \delta_{ss'} Y_s^2, \cm
         Y_s^2 =
               d(I)\bigl[1- d(I)/(D-2) \bigr] + \olam^2_{s} >0
\eeq
     where $\olam^2_s = \sum_a\lambda^2_{sa}$.
     Then the functions $y_s(u)$ obey the decoupled Liouville equations
     $\ddot y_s = \ep_s Q^2_s Y^2_s \e^{2y_s}$, whence
\beq                                                    \label{3.23}
     \e^{-2y_s(u)} = \vars{
              Q^2_s Y^2_s\, s^2(h_s,\ u+u_s),            & \ep_s = 1,\yy
              Q^2_s Y^2_s h_s^{-2} \cosh^2 [h_s(u+u_s)], & \ep_s = -1,
                            \quad  h_s > 0,}
\eeq
     where $h_s$ and $u_s$ are integration constants and the function
     $s(.,.)$ has been defined in (\ref{3.8}). For the sought-for functions
     $x^A(u)$ and the ``conserved energy'' $E$ we then obtain:
\bear                                                        \label{3.24}
     x^A(u) \eql \sums \frac{Y_s{}^A}{\Ysq} y_s(u) + c^A u + \uc^A,
 \\
     E \eql \sums \frac{h_s^2\sign h_s}{\Ysq} + \vec c\,{}^2 \label{3.29}
                     = \frac{d_0}{d_0-1} \ k^2 \sign k,
\ear
     where the vectors of integration constants $\vec c$ and $\vec\uc$ are
     orthogonal to each $\vY_s$: \ $c^A Y_{s,A} = \uc^A Y_{s,A} = 0$.

     Although many other solutions are known \cite{IM2,IM5,bobs}, their
     physical properties turn out to be quite similar to those of
     the present solutions for orthogonal systems (OS)
     of vectors $\vY_s$ \cite{bobs,br-jmp}.

     Black holes (BHs) are distinguished among other \sph\ solutions by the
     requirement that there should be horizons rather than singularities
     in \Mext\ at $u=\umx$. This leads to constraints upon the input
     and integration constants. The above OS solutions describe BHs if
\bear
      h_{s} = k >0, \cm \forall\ s; \cm\cm
         c^A = k \sums Y_{s}^{-2} Y_{s}{}^A - k \delta^A_1,    \label{5.4}
\ear
     where $A=1$ corresponds to $i=1$ (time). The constraint (\ref{3.29})
     then holds automatically. The value $u = \umx = \infty$ corresponds to
     the horizon. The no-hair theorem of Ref.\,\cite{br-jmp} states that
     BHs are incompatible with quasiscalar $F$-forms, so that all $\ep_s=1$.

     Under the asymptotic conditions $\varphi^a \to 0$, $\beta^i \to 0$
     as $u\to 0$, after the transformation
\bearr
     \e^{-2ku} = 1 -\frac{2k}{\od r^\od},\cm \od := d_0-1  \label{5.5}
\ear
     the metric (\ref{3.10}) for BHs and the corresponding scalar fields
     may be written as
\bearr
     ds_D^2 =
     \biggl(\prod_{s} H_{s}^{A_{s}}\biggr) \Biggl[-dt^2
          \biggl(1-\frac{2k}{\od r^\od}\biggr)\prod_{s} H_{s}^{-2/Y_s^2}
     + \biggl(\frac{dr^2}{1-2k/(\od r^\od)} + r^2 d\Omega^2\biggr)
          + \sum_{i=2}^{n} ds_i^2
                       \prod_s H_{s}^{A_{s}^i}\Biggr];
\nnnv                           \label{5.7}
    \inch  A_{s} := \frac 2{\Ysq}\frac{d(I_s)}{D-2};
    \cm
             A_{s}^i := -\frac{2}{\Ysq} \delta_{iI_s};
\yyy
     \varphi^a = -\sums \frac{\lambda_{sa}}{\Ysq} \ln H_s,    \label{phi}
\ear
     where $H_{s}$  are harmonic functions in $\R_+ \times \S^{d_0}$:
\beq                                                            \label{5.8}
     H_{s} (r) =  1 + P_{s}/(\od r^\od), \cm
                    P_s := \sqrt{k^2 + \Qsq \Ysq} - k.
\eeq

     The subfamily (\ref{5.4}), (\ref{5.7})--(\ref{5.8}) exhausts all OS BH
     solutions with $k > 0$ (non-extremal BHs). Extremal BHs, corresponding
     to minimum mass for given charges (the so-called BPS limit), are
     obtained either in the limit $k \to 0$, or directly from
     (\ref{3.23})--(\ref{3.29}) under the conditions $h_s = k = c^A =0$.
     The only independent integration constants in the BH solutions
     are $k$, related to the observed mass (see below), and the brane
     charges $Q_s$.

     Other families of solutions, mentioned previously,
     also contain BH subfamilies. The most general BH solutions
     are considered in Ref.\,\cite {im9910}.

\subsection {Post-Newtonian parameters and other observables}

     One cannot exclude that real astrophysical objects (stars, galaxies,
     quasars, black holes) are essentially multidimensional objects,
     whose structure is affected by charged \branes. (It is then unnecessary
     to assume that the antisymmetric form fields are directly observable,
     though one of them may manifest itself as the electromagnetic field.)

     The {\bf post-Newtonian} (PN) (weak gravity, slow motion) approximation
     of multidimensional solutions then determines the predictions of the
     classical gravitational effects: gravitational redshift, light
     deflection, perihelion advance and time delay (see \cite{will,damr}).

     For \sph\ configurations, a standard form of the PN metric
     uses the Eddington parameters $\beta$ and $\gamma$ in isotropic
     coordinates, in which the spatial part is conformally flat \cite{will}:
\beq
      ds_{\rm PN}^2 =                                           \label{gPN}
            - (1 -2V + 2\beta V^2) dt^2
             + (1+ 2 \gamma V) (d\rho^2 + \rho^2 d\Omega^2)
\eeq
     where $d\Omega^2$ is the metric on $\S^2$, $V = GM/\rho$ the
     Newtonian potential, $G$ the Newtonian gravitational constant and
     $M$ the active gravitating mass.

     Observations in the Solar system lead to
     tight constraints on the Eddington parameters \cite{damr}:
\bear                                           \label{obs_}
    \gamma = 0.99984 \pm 0.0003, \cm
    \beta = 0.9998 \pm 0.0006.
\ear
     The first restriction results from over VLBI observations
     \cite{Reas}, the second one from the $\gamma$ data and an analysis of
     lunar laser ranging data \cite{Nor,Dik}.

     For a theory under consideration, the metric (\ref{gPN}) should be
     identified with the asymptotics of the 4D metric in the
     observational CF. Preserving its choice yet undetermined, we can write
     according to (\ref{3.10}) with $d_0=2$:
\beq
     ds^*_4= \e^{2f(u)}                                          \label{g4}
         \biggl\{-\e^{2\beta^1} dt^2
                + \frac{\e^{-2\sigma_1}}{s^2(k,u)} \biggl[
            \frac{du^2}{s^2(k,u)} + d\Omega^2\biggr]\biggr\}
\eeq
     where $f(u)$ is an arbitrary function of $u$, normalized for
     convenience to $f(0) =0$ (not to be confused with $f(\phi)$
     that appeared in \sect 2 and 3). Recall that by our notations
     $\sigma_1 = \beta^1 + \sigma_2$, the function $s(k,u)$ is defined in
     \eq (\ref{3.8}), and spatial infinity takes place at $u=0$.

     Passing to isotropic coordinates in (\ref{g4}), one finds that
     $u = 1/\rho$ up to cubic terms in $1/\rho$, and the decomposition in
     $1/\rho$ up to $O(\rho^{-2})$, needed for comparison with (\ref{gPN}),
     precisely coincides with the $u$-decomposition near $u=0$.

     Using this circumstance, it is easy to obtain for the mass
     and the Eddington parameters corresponding to (\ref{g4}):
\beq
     GM = -\bopr - f';                                       \label{Ed-f}
            \qquad
     \beta = 1 + \Half\,\frac{\boprr + f''}{(GM)^2},
            \qquad
     \gamma = 1 + \frac{2f' - \sigma_2'}{GM},
\eeq
     where $f' = df/du\Big|_{u=0}$ and similarly for other functions. The
     expressions (\ref{Ed-f}) are quite general and are applicable to
     asymptotically flat, \ssph\ solution of any theory where the
     energy-momentum tensor has the property $T^u_u + T^\theta_\theta =0$,
     which leads to the metric (\ref{3.10}) and, in particular, to the
     above solutions of the theory (\ref{act-F}).

     Two special choices of $f(u)$ can be distinguished. First, if, for some
     reasons, the 4D Einstein frame is chosen as the observational one,
     then, according to \eq (\ref{conf}) with $d_0=2$, we have
\beq
    f = f^{\rm E} = \sigma_2/2.                        \label{4-E}
\eeq
     Second, let us try to add the matter action in (\ref{act-F}) simply
     as $\const\cdot \int L_m \sqrt{{}^D g}\, d^D x$, i.e., like the
     fermionic
     terms in the effective action in the field limit of string theory
     (\cite{GSW}, \eq (13.1.49)]. In the observational
     frame with the metric $g^*_{\mu\nu}$ the matter action should read
     simply $\int d^4 x \sqrt{g^*} \,L_m$. Identifying them, we obtain
     \cite{br95-2,pn} $g^*\mn  = \e^{\sigma_2/2}g\mn$, whence
\beq
           f = f^* = \sigma_2/4.                              \label{g*}
\eeq

     The parameter $\beta$ can be calculated using (\ref{Ed-f}) {\it directly
     from the equations of motion (\ref{eqm}), without solving them\/}.
     This is true for any function $f$ of the form $f = \vec F \vec x$ where
     $\vec F \in \V$ is a constant vector (i.e., $f$ is a linear combination
     of $\beta^i$ and the scalar fields $\varphi^a$):
\beq
     \beta - 1 = \frac{1}{2(GM)^2} \sums                   \label{be-f}
             \ep_s Q_s^2 (Y_s^1 + \vec F \vY_s) \e^{2y_s(0)}.
\eeq

     Explicit expressions for $M$ and $\gamma$ require
     knowledge of the solutions' asymptotic form. However,
     there is an exception: due to (\ref{4-E}),
     in the 4D Einstein frame $\gamma = 1$, precisely as in GR, for all
     \brane\ solutions in the general model (\ref{act-F}).

     Let us also present the quantities $\bopr$ and $\sigma'_2$, needed for
     finding $GM$ and $\gamma$, for OS BH solutions:
\beq
     \bopr = -k - \sums P_s \frac{1-b_s}{\Ysq}, \cm
     \sigma'_2 = -\sums \frac{1 - 2b_s}{\Ysq}                 \label{parBH}
\eeq
     where $b_s=d(I_s)/(D-2)$.

     Some general observations can be made from the above relations
     \cite{pn}:
\begin{itemize}
\item
     The expressions for $\beta$ depend on the input constants $D$, $d(I_s)$
     (hence on \brane\ dimensions), on the mass $M$ and on the charges
     $Q_s$. For given $M$, they are independent of other integration
     constants, emerging in the solution of the Toda system (\ref{eqm}), and
     also of \brane\ intersection dimensions, since they are obtained
     directly from \eqs (\ref{eqm}) \cite{pn}. This means, in
     particular, that $\beta$ is the same for BH and non-BH configurations
     with the same set of input parameters, mass and charges.
\item
     All \branes\ give positive contributions to $\beta$ in both frames
     (\ref{4-E}) and (\ref{g*}), which leads to a general restriction on the
     charges $Q_s$ for given mass and input parameters.
\item
     The expressions for $\gamma$ depend, in general, on the integration
     constants $h_s$ and $c^i$ emerging from solving \eqs (\ref{eqm}). For
     BH solutions these constants are expressed in terms of $k$ and the input
     parameters, so both $\beta$ and $\gamma$ depend on the mass, charges
     and input parameters.
\item
     In the 4-E frame, one always has $\gamma=1$. The same is true for some
     BH solutions in all frames with $f=N \sigma_2,\ N=\const$ \cite{pn}.
\end{itemize}

\noi{\bf BH temperature.}
     BHs are, like nothing else, strong-field gravitational objects, while
     the PN parameters only describe their far neighbourhood. An important
     observable characteristic of their strong-field behaviour
     is the Hawking temperature $\TH$. One can show \cite{pn} that this
     quantity is {\it CF-independent\/}, at least if conformal factors that
     connect different frames are regular on the horizon. The conformal
     invariance of $\TH$ was also discussed in another context in
     Ref.\,\cite{JaK}.

     In particular, for the above OS BH solutions one obtains \cite{bim97}
\beq
     \TH = \frac{1}{8\pi k k_{\rm B}}                      \label{THBOS}
        \prod_s \biggl(\frac{2k}{2k+P_s}\biggr)^{1/\Ysq}.
\eeq
     where $k_{\rm B}$ is the Bolzmann constant.

     The physical meaning of $\TH$ is related to quantum evaporation, a
     process to be considered in the fundamental frame, while the produced
     particles are assumed to be observed at flat infinity, where relevant
     CFs do not differ. Therefore $\TH$ should be CF-independent,
     and this property is obtained ``by construction'' \cite{pn}.

     All this is true for $\TH$ in terms of the integration constant $k$ and
     the charges $Q_s$. However, the observed mass $M$ as a function of the
     same quantities is frame-dependent, see (\ref{Ed-f}). Therefore $\TH$
     as a function of $M$ and $Q_s$ is frame-dependent as well.

\medskip\noi
{\bf Coulomb law violation} is one of specific potentially observable effects
     of extra dimensions. Suppose in (\ref{act-D}), (\ref{dsDn}) $d_0=2$
     and let us try to describe the electrostatic field of a \sph\ source
     by a term $F^2 \e^{2\olam\ophi}$ in the action
     (\ref{act-F}), corresponding to a true electric $m$-form $\Fei$ with
     a certain set $I$ containing 1, that is, $I = {1} \cup J$,
     $J \subset \{2,\ldots, n\}$.

     Then the modified Coulomb law in any CF with the metric (\ref{g4})
     can be written as follows \cite{pn}:
\beq                                                             \label{Ef}
     E = = (|Q|/r^2)\e^{-2\olam\ophi + \sigma(J) - \sigma(\oJ)},
\eeq
     where $E$ is the observable electric field strength, $r$ is the
     observable radius of coordinate spheres, the notations
     (\ref{sigma}) are used and $\oJ = \{2,\ldots, n\}\setminus J$.

     Deviations from the conventional Coulomb law are evidently both due
     to extra dimensions (and depend on the $F$-form structure) and due
     to interaction of $F_s$ with the scalar fields. This relation
     (generalizing
     the one obtained in Ref.\,\cite{bm-ann} in the framework of dilaton
     gravity) is valid for an arbitrary metric of the form (\ref{dsDn})
     ($d_0=2$) and does not depend on whether or not this $F$-form takes
     part in the formation of the gravitational field.

     \eq (\ref{Ef}) is exact and --- which is remarkable --- it is {\it
     CF-independent}. This is an evident manifestation of the conformal
     invariance of the electromagnetic field in $\Mext$ even in the present
     generalized framework.

\section{Concluding remarks}

     We have seen that the properties of theoretical models look
     drastically different when taken in different CFs. This once again
     stresses the necessity of a careful reasoning for a particular choice
     of a CF. It even may happen (though seems unlikely) that there is
     an unobservable part of the Universe, separated from us by a
     singularity in the observational CF which is converted to a regular
     surface (\Str) after passing to a fundamental frame. A similar thing
     may happen to a cosmological singularity: in some theories it can
     correspond to a regular bounce in a fundamental frame.

     We have obtained expressions for the Eddington PN parameters $\beta$
     and $\gamma$ for a wide range of \ssph\ solutions of multidimensional
     gravity with the general string-inspired action (\ref{act-F}).
     The experimental limits (\ref{obs_}) on $\beta$ and $\gamma$
     constrain certain combinations of the solution parameters. This,
     however, concerns only the particular physical system for which the
     measurements are carried out, in our case, the Sun's gravitational
     field. The main feature of the expressions for $\beta$ and $\gamma$ is
     their dependence not only on the theory (the constants entering into
     the action), but on the particular solution (integration constants).
     The PN parameters thus can be different for different self-gravitating
     systems, and not only, say, for stars and \bhs, but even for different
     stars if we try to describe their external fields in models like
     (\ref{act-F}).

     A feature of interest is the universal prediction of $\beta > 1$ in
     (\ref{be-f}) for both frames (\ref{4-E}) and (\ref{g*}).
     The predicted deviations of $\gamma$ from unity may be of any sign and
     depend on many integration constants. If, however, the 4-dimensional
     Einstein frame is adopted as the observational one, we have a universal
     result $\gamma=1$ for all \ssph\ solutions of the theory (\ref{act-F}).

     The BH temperature $\TH$ also carries information about the
     space-time structure, encoded in $\Ysq$. Being a universal parameter
     of a given solution to the field equations, $\TH$ as a function of the
     observable BH mass and charges is still CF-dependent due
     to different expressions for the mass $M$ in different frames.

     One more evident consequence of extra dimensions is the Coulomb
     law violation, caused by a modification of the conventional Gauss
     theorem and also by scalar-electromagnetic interaction. A remarkable
     property of the modified Coulomb law is its CF independence for a
     given \ssph\ metric.


\def\Jl#1#2{{\it #1\/} {\bf #2},\ }

\def\CQG#1 {\Jl{Class. Quantum Grav.}{#1}}
\def\DAN#1 {\Jl{Dokl. AN SSSR}{#1}}
\def\GC#1 {\Jl{Grav. \& Cosmol.}{#1}}
\def\GRG#1 {\Jl{Gen. Rel. Grav.}{#1}}
\def\JETF#1 {\Jl{Zh. Eksp. Teor. Fiz.}{#1}}
\def\JMP#1 {\Jl{J. Math. Phys.}{#1}}
\def\NP#1 {\Jl{Nucl. Phys.}{#1}}
\def\NPB#1 {\Jl{Nucl. Phys.}{B\ #1}}
\def\PLA#1 {\Jl{Phys. Lett.}{#1A}}
\def\PLB#1 {\Jl{Phys. Lett.}{#1B}}
\def\PRD#1 {\Jl{Phys. Rev.}{D\ #1}}
\def\PRL#1 {\Jl{Phys. Rev. Lett.}{#1}}

\end{document}